\documentstyle[psfig,proceedings]{crckapb}
\def\lisix{$^6$Li\,}
\def\liseven{$^7$Li\,}
\def\Msun{$M_{\odot}$\,}

\begin{opening}
\title{SURVIVAL OF \lisix \, AND \liseven \, IN METAL-POOR STARS}
\author{R. Cayrel$^1$}
\institute{$^1$Observatoire de Paris \\ 61, av. de l'Observatoire F-75014 Paris, France}
\author{Y. Lebreton$^1$}
\institute{$^1$ Observatoire de Paris, Section de Meudon \\
 F-92195 Meudon Cedex, France}
\author{P. Morel$^2$}
\institute{$^2$ Observatoire de la C\^ote d'Azur \\
BP 229 \\ F-06304 Nice Cedex 4 }
\end{opening}
\begin{document}
\begin{abstract}
The relationship between the depletions of \lisix and \liseven is studied 
for two models of lithium burning, below the convective zone. The parameters
of the depletion models are submitted to the constraint that the slope of
the \liseven theoretical depletion curve agrees with the slope of the
observed depletion curve, for cool subdwarfs. Other less restrictive models
are also considered. \\
In all cases, a \lisix depletion less than 0.5 dex implies a \liseven depletion
less than 0.1 dex. With the constraint on the slope of the \liseven curve, the depletion 
of \liseven for the same depletion of \lisix is below 0.05 dex.
 
The still unsolved problem for the true \liseven abundance in subdwarfs is the possible
influence
of temperature inhomogeneities, raised by Kurucz, subsequently shown to be small in the
solar case, but not yet computed with the inclusion of departure from LTE for 
metal-poor stars. 

\end{abstract}

\section{Introduction} Lithium is one of the great fossil records available
in  our local stellar environment. It was discovered by the Spites \cite{SS82}
in subdwarfs.

Apparently, the \liseven /H ratio, has survived all the way from 
100 seconds after the Big Bang, to the birth of the first stars, and, more or less,
subsequently
in the atmospheres of these  stars, in which we observe it, now, 13 Gyr later.
The value of this ratio, based on the analysis of 40 metal-poor
stars (Bonifacio \& Molaro \cite{BM97}), is about 1.6$\times 10^{-10}$ (by number,
or $1.1\times 10^{-9}$ by mass),  close to
the Standard Big Bang Nucleosynthesis prediction (Schramm \& Turner \cite{ST98}),
for the $\eta $ ratio (baryons/photons) set by the other products of the SBBN.

The debated question is that of the survival of \liseven in the atmospheres
of the stars, during the 12 or 13 Gyr they have spent on the main sequence.
Two processes can alter the initial lithium content: gravitional settlling at the bottom
of the convective zone (CZ hereafter) (Vauclair \& Charbonnel \cite{VC98}, and poster 
at this meeting), and mixing of the matter in the CZ with deeper layers 
, in which the temperature allows nuclear burning of \liseven, (Schatzman (1977),  
Pinsonneault et al. \cite{PDD92}, Charbonnel et al. \cite{CVZ92}, Montalban \& Schatzman
\cite{MS96}). Furthermore, Kurucz (1995) has suggested that the abundance of lithium
determined with plane-parallel atmospheres might be strongly in error, due to  effects
of temperature inhomogeneities in the convective zone, closer to the 
surface in metal-poor stars than in solar composition stars.

We shall discuss here only the main process, the burning process by  mixing of the CZ 
with deeper layers, 
clearly acting on metal-poor stars of effective temperature below 5700 K (see fig. 1),
and responsible  of a depletion of \liseven by a factor of 200 in the Sun. 

As \lisix has been also observed in two turnoff metal-poor stars, HD 84937 and HD 338529, 
and burns at a lower temperature than \liseven, it is  interesting to see
what constraints the observation of \lisix brings on the maximum depletion of 
\liseven .  
 
\section{Burning of \lisix and \liseven during pre-main-sequence}

The burning of  these elements during pre-main-sequence has been computed using
the CESAM code (Morel \cite{M97}. Opacities are from Rogers \& Iglesias \cite{RI92},
Rogers et al. \cite{RSI96}, Alexander \& Fergusson \cite{AF94}, or Kurucz \cite{K95}
The EFF equation of state is used (Eggelton et al. \cite{EFF73}). The results are 
given  in columns 3 and 6 of table 1.
\begin{table}
\caption[]{Depletion of \lisix and \liseven, in dex, during the PMS and MS phases
( values between parentheses are extrapolated and less reliable)}
\begin{flushleft}
\begin{tabular}{llllllll}
\hline
$ M $ & [Fe/H] & \lisix (PMS) &\lisix (MS) & \lisix (tot)\hspace{2mm}&\liseven (PMS) & \liseven (MS)  & \liseven (tot)\\
\Msun & dex & dex & dex & dex & dex & dex & dex \\ 
\hline
0.85  &  -1.5  &  0.20    & 0.48  & (.74) & 0.0 & 0.06 & 0.06 \\
0.85  &  -2.0 &   0.11 & 0.01  & 0.12   &  0.0 & 0.00  & 0.00  \\
0.80  &  -1.5 &  0.411  &1.06  & 1.47  &  0.01 & 0.37  & 0.38 \\
0.80  & -2.0  &  0.32  & .52 & 0.84  &  0.00 & 0.06  & 0.06 \\ 
0.75  & -1.5  &   0.92 &(2.3) & (3.22) & 0.01 & 0.87  & 0.88 \\
0.75  &  -2.0 &  0.81  & 1.23 & 2.04 &  0.01  & 0.56 &  0.57 \\
\hline
\end{tabular}
\end{flushleft}
\end{table}

It must be noted that there is in all cases some depletion of \lisix during
the PMS, and that this depletion reaches 0.4 dex at metallicity -1.5 for a
mass of 0.80 \Msun . PMS depletion of \liseven is negligible in the range of masses
considered here. Depletions are counted as positive for a reduction in lithium abundance
(e.g. a depletion of 1.0 dex means that the present abundance is one tenth of 
the initial abundance). 

\section{Burning of \lisix  and \liseven during MS lifetime and after}
\subsection{predictions based on a simple circulation model}
We shall use first a very simple model of destruction  based on the fact that 
the rate of destruction of \lisix and \liseven varies as a function of depth
almost as a step function.
 In the convective zone itself, and just below, the destruction rate 
is so small that no depletion of these elements occur in the time available (12  to 14 Gyr).
On a short length, of about one hundredth of the radius of the star, the rate increases
to such a large value that the element is burnt everywhere below, in a time  much smaller than
the age of the object. The depletion in the well-mixed convective zone depends only
upon its exchange of matter with the zone in which  the element
is  totally burnt. The corresponding level is of course deeper for \liseven than
for \lisix. For example, for a star of mass 0.80 \Msun and [Fe/H]=-2.0, the 
transition occurs for \liseven at a depth of 0.155 below the bottom of the CZ, in units of the 
stellar radius,  but only at a depth of 0.10 for \lisix . Under the assumption
that the exchange of matter is driven by convective plumes penetrating the radiative
zone, as suggested by hydrodynamical simulations (Stein \& Nordlund \cite{SN89}, Rieutord \&
Zahn \cite{RZ95}) , the critical parameter for the exchange of matter between the CZ 
and the burning zone is of course the separation  between the two levels.
\begin{figure}
\centerline{
\psfig{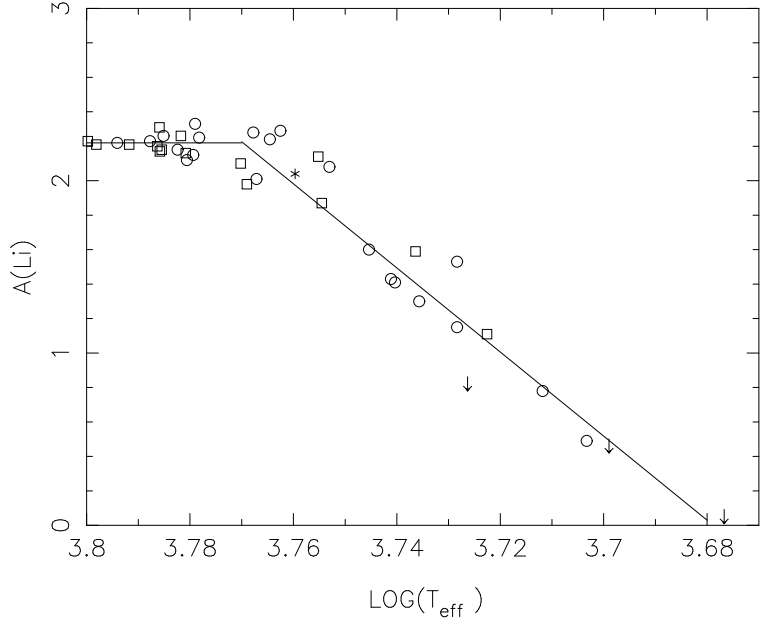}}
\caption[]{Abundance of Li as a function of $\log T_{eff} $. The sources are 
Bonifacio \& Molaro \cite{BM97} and Ryan \& Deliyannis \cite{RD98}. The abundances
are given in the usual logarithmic scale, in which the abundance of hydrogen
is 12.0. These abundances are the total abundance of Li, in practice, they are
abundances of  \liseven , \lisix representing at the most 5 per cent of the
abundance of \liseven , and most of the time much less.}
\end{figure}

On fig. 1 we have plotted depletions
of  \liseven according to two recents papers (Bonifacio \& Molaro \cite{BM97}, Ryan \& 
Delyiannis \cite{RD98}), as a function of $\log(T_{eff})$. We have also computed the
distance $d$ of the top of the burning zone to the bottom of the convective zone,
on the main sequence, with the CESAM code, for a grid of models , with masses
0.70[0.05]0.85 \Msun ( the value between brackets is the step),  metallicities [Fe/H]= -2.0[0.5]-1.0  an helium abundance
near Y$_p$ = 0.23, and an enhancement of $\alpha$-elements by 0.4 dex. As $T_{eff}$
is a result of the computation, this allows to redraw the depletion of \liseven as a function of 
$d$. In this transformation we have allowed a shift of 0.01 between the theoretical
$\log(T_{eff})$ and the observed  $\log(T_{eff})$,  as shown necessary in Cayrel et al.
\cite{CLPT97}.

\begin{figure}
\centerline{
\psfig{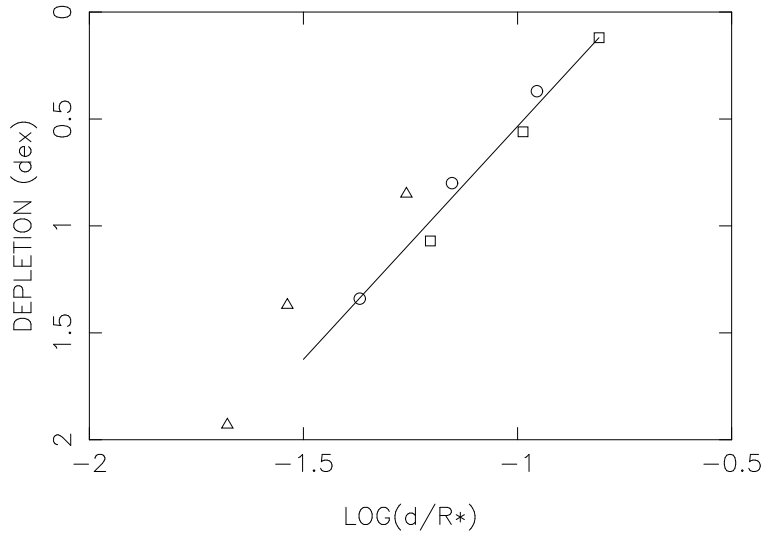}}
\caption[]{Depletion of  \liseven as a function of the distance $d$ between 
the bottom of the convective zone and the top of the burning region. The
unit of length is the radius of the star. The symbols are : squares: 
metallicity [Fe/H]=-2.0; circles: [Fe/H]= -1.5; triangles: [Fe/H]= -1.0.
The straight line is a global fit for the two lowest metallicities.}
\end{figure}  
 The result, shown in fig. 2, confirms the idea that $d$ is a key 
parameter for the control of depletion by burning. The two straight lines
for the two metallicities [Fe/H] = -1.5 and -2.0 can be approximated by the 
single relation:
\begin{equation}
 D(^7{\rm Li} ) = 0.,\  {\rm for} \  d > 0.16R_* \label{eq:1}    
\end{equation}
\begin{equation}
 D(^7{\rm Li}) =  -2.7(\log(d/R_* ) + 0.83)\ \  {\rm for}\  d < 0.16R_* \label{eq:2}
 \end{equation}  

The physical interpretation of this is that if the rate of exchange of matter between
the convective zone and the burning zone  is   $a_{^7{\rm Li}}$, in fraction of mass of the CZ
per Gyr, the  depletion factor of \liseven  after  a time $t$ will be: 

$$R = \exp(-a_{^7{\rm Li}} t)$$

\noindent as the matter returning from the burning zone has a zero-content in \liseven.
In decimal exponent the depletion becomes $ D=0.4343*a_{^7{\rm Li} }t $
or, for an age of 13 Gyr: $ D=5.65\times a_{^7{\rm Li}}$. For a depletion of 0.5 dex
this gives $ a_{^7{\rm Li}} = 0.088 $, or a circulation rate a little below
one tenth of the mass of the CZ per Gyr.
Because the circulation rate as a function of depth is independent of the 
element considered, formulae (1) and (2) are applicable as well to \lisix,
at the only condition that $d$ refers to the depth corresponding to
the top of the burning zone of \lisix and not to the top of the burning zone
of \liseven. This is how we have computed the predicted values for \lisix,
actually derived from the observed depletions of \liseven for the same 
distance between the bottom of the CZ and the top of the corresponding 
burning zone. The depletions of \lisix and \liseven obtained are plotted in
fig. 3, as open circles. Up to a depletion of 1.0 for \lisix the 
depletion of \liseven remains below 0.1 dex. The main cause of this 
is the strong depletion of \lisix in the PMS phase. 

\begin{figure}
\centerline{
\psfig{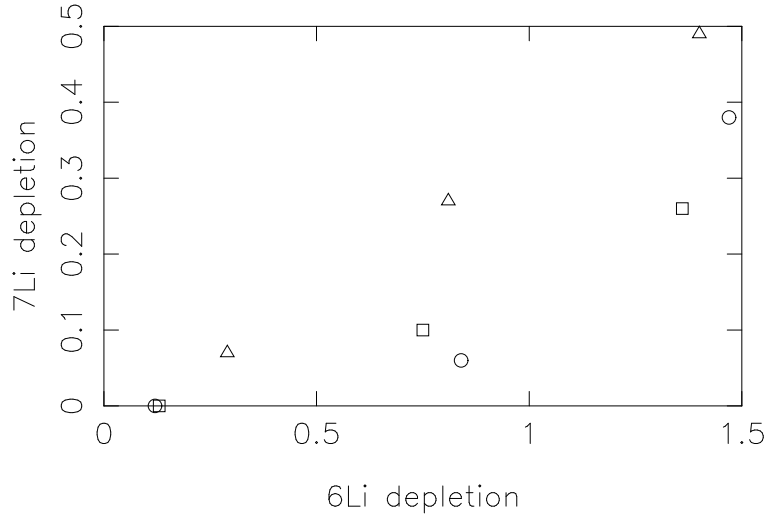}}
 \caption[]{Relationship between the respective depletions of \lisix and
\liseven . The open circles are the results of the simple circulation model
 of section 3.1. The squares are the results obtained with a mixing by random-walk,
 with a diffusion coefficient adjusted to reproduce the depletion curve
 of \liseven in fig.1. The triangles are for a model with a diffusion coefficient
 independent of depth.}
\end{figure}

\subsection{predictions based on a random-motion mixing model}

The first model of mixing was proposed by E. Schatzman \cite{S77}, and is
generally described as mixing by "turbulence", or "diffusion". We shall 
adopt here such a model, phenomenogically described by random-motions 
characterized locally by the value of a diffusion coefficient, function of
depth. If the source of the stirring is convection, one expects the diffusion
coefficient to decrease stronly with depth below the bottom of the CZ (see
Montalban \& Schatzman \cite{MS96}). Actually we can adjust the value at some arbitrary 
depth and the scale-height of an assumed exponential decline of the diffusion
coefficient in such a way that the abundance of \liseven follows the behaviour
seen in fig. 1. In fig. 3 the squares show the relationship between the two
depletions for this model. The conclusions are the same as for the first model.
An extremely conservative bound for the depletion of \liseven at a given 
depletion of \lisix is given by a model for which the diffusion coefficient 
is constant with depth (triangles in fig. 3). Even in that case the depletion
of \liseven is les than 0.1 up to depletions of \lisix of about 0.4. For this last model
the constraint of reproducing the slope of the depletion curve of fig. 1 is
not fullfilled.

 \section{Conclusions}
 
Models of destruction by burning , reproducing the slope of the depletion curve
of \liseven, in unevolved metal-poor stars cooler than 5800 K, show a fairly
robust ratio of the depletions of \lisix and \liseven in the range of interest.
The depletion of \liseven remains below 0.1 dex , as soon as the depletion of
\lisix is below the range 0.5--1.0. 
 This result is valid both for a random motion
type of mixing or for a stationary circulation type of mixing. A large fraction of \lisix
burning occurs during the pre-main sequence. 

The burning of \lisix leads to an important depletion as soon as one of these two 
conditions is  met: either a mass below 0.80 M$_{\odot}$, or [Fe/H] $> -1.5$.
This helps to understand why \lisix has been found so far only in TO or subgiants more
metal-poor than -1.5.

Although there are some indications that other effects, as gravitational settling, and
temperature inhomogeneities in convective elements are not very important for \liseven,
a global and consistent treatment including microscopic diffusion, mixing, mass loss and temperature
inhomogeneities  is a must for the coming years.

\begin{acknowledgements}
Part of this work has been performed using the computing facilities 
provided by the OCA program
``Simulations Interactives et Visualisation en Astronomie et M\'ecanique 
(SIVAM)''.
\end{acknowledgements}

\end{document}